\documentclass{optica-article}
\makeatletter
\newif\ifosa@openaccess
\makeatother
\journal{opticajournal} 

\articletype{Research Article}

\usepackage{lineno}

\usepackage{braket}
\usepackage{siunitx}
\usepackage{amsmath}
\usepackage{printlen}
\usepackage{graphicx}
\usepackage{float}
\uselengthunit{mm}
\usepackage{hyperref}
\usepackage{booktabs}
\DeclareSIUnit\counts{cps}

\begin{document}

\title{Quantum teleportation with photons from remote quantum dots emitting at telecom C-band}

\author{Raphael Joos,\authormark{1,*} Michelle Pfahl,\authormark{1} Ponraj Vijayan,\authormark{1} Michal Vyvlecka,\authormark{1} Michael Jetter,\authormark{1} Simone L. Portalupi,\authormark{1} and Peter Michler\authormark{1} }

\address{\authormark{1}Institut für Halbleiteroptik und Funktionelle Grenzflächen (IHFG), Center for Integrated Quantum Science and Technology (IQ$^{ST}$) and SCoPE, University of Stuttgart, Allmandring 3, 70569 Stuttgart, Germany\\}

\email{\authormark{*}r.joos@ihfg.uni-stuttgart.de} 


\begin{abstract*} 
Employing quantum teleportation to increase the communication distance in quantum channels requires distinct quantum light sources, preferably operating at telecommunication wavelength. Here we utilize two semiconductor In(Ga)As quantum dots, strain engineered to emit in the telecom C-band, as sources of single- and entangled-photons, respectively. This enables a full-photonic quantum teleportation experiment, achieving a maximum teleportation fidelity of \SI{92.7\pm0.5}{\percent}. Owing to state-of-the-art source brightness, and high time resolution, it is here possible to employ the three-photon coincidences to fully quantify the time-dependence of the process. In this way, we highlight current limitations and implement an optimization in the data-processing which leads to the maximization of achievable teleportation fidelity and coincidence rates. These results constitute a step forward in practical implementations of quantum teleportation in future fiber-based quantum networks.
\end{abstract*}

\section{Introduction}
Establishing a global quantum internet is a highly rewarding long-term goal, but its realization presents significant technological and scientific challenges. An important building block for such long range networks is the quantum repeater~\cite{VanLoock2020, Azuma2023}, which will drastically increase the communication distance between nodes. A first step towards quantum repeaters is quantum teleportation between remote network nodes~\cite{Bennett1993}. Since its first experimental demonstration in $1997$~\cite{Bouwmeester1998}, quantum teleportation has evolved from an intriguing fundamental physics experiment into a promising tool for practical applications~\cite{Ma2012, Bussieres2014, Hermans2022}. Key elements for such measurements are efficient sources of quantum light, favorably at telecommunication wavelengths for diminishing losses when propagating through silica fibers. Proof-of-principle experiments of quantum teleportation have been demonstrated with parametric processes~\cite{Bouwmeester1998}, and various single emitters, from atoms and ions~\cite{Barrett2004,Riebe2004,Olmschenk2009}, to color centers in diamond~\cite{Pfaff2014}. Recently, research on quantum dot (QD)-based quantum teleportation has progressed from seminal studies~\cite{Gao2013,Reindl2018,Anderson2020,BassoBasset2021} to experiments demonstrating full-photonic teleportation with two distinct sources:
This was demonstrated using GaAs QDs embedded in optical resonators and operating at near-infrared (NIR) wavelengths~\cite{Laneve2025}. At the same time, another experiment utilized similar QDs in combination with quantum frequency conversion to perform the Bell state measurement (BSM) at telecom wavelength~\cite{Strobel2025}.\\
Here, we demonstrate successful quantum teleportation using two quantum dots that emit directly within the technologically relevant telecom C-band. For this purpose, we utilized our state-of-the-art source of entangled photon pairs~\cite{Joos2026} operated within our high-efficiency stand-alone setup~\cite{Vyvlecka2026} in combination with a second sample serving as source of single photons. To enable interference within the BSM, the single-photon source is spectrally tuned using a piezoelectric actuator~\cite{Trotta2016}. The teleportation process is investigated in terms of its full time-dependence on the detection time separation of the three photons. We report a maximum achievable teleportation fidelity of \SI{92.7\pm0.5}{\percent}. Furthermore, owing to the time-resolved measurements of the three-photon coincidences, it is possible to quantify the impact of source imperfections. This allows for a novel optimization of data-processing, which leads to maximization of achievable teleportation fidelity and coincidence rates. Based on the figures-of-merit, teleportation fidelity and coincidence rate, this work surpasses previous state-of-the-art implementations of quantum teleportation with remote QD~\cite{Laneve2025,Strobel2025} and compares well with other photonic quantum teleportation platforms~\cite{Olmschenk2009, Nolleke2013, Ito2025, Akin2025, Peng2025}.

\section{Theory}
We perform quantum teleportation of an arbitrary input state of photon 1
\begin{equation}
	\ket{\Psi_1^\text{in}} = \alpha\ket{H_1} + \beta\ket{V_1}, \quad |\alpha|^2 + |\beta|^2 = 1
\end{equation}
with an entangled state $\ket{\Psi_{23}^\text{ent}(\tau_{23})}$ of the photons 2,~3 emitted by the biexciton-exciton-cascade of a semiconductor QD. Due to a non-zero fine-structure splitting $FSS$, the entangled state exhibits a time-evolving phase dependent on the time delay $\tau_{23}$ between the biexciton and exciton photons of the cascade
\begin{equation}
	\ket{\Psi_{23}^\text{ent}(\tau_{23})} = \frac{1}{\sqrt{2}}\left(\ket{H_2H_3}+\text{e}^{i\frac{FSS}{\hbar}\tau_{23}}\ket{V_2V_3}\right).
\end{equation}
After the interference of photons 1 and 2 on a balanced beam splitter (BS), the three-photon-state can be written as~\cite{Reindl2018}
\begin{equation}
	\begin{split}
		\ket{\Psi_{123}(\tau_{23})} &= \ket{\Psi_1^\text{in}}\otimes\ket{\Psi_{23}^\text{ent}(\tau_{23})} =  \left(\alpha\ket{H_1} + \beta\ket{V_1}\right) \otimes\frac{1}{\sqrt{2}}\left(\ket{H_2H_3}+\text{e}^{i\frac{FSS}{\hbar}\tau_{23}}\ket{V_2V_3}\right)\\
		&= \frac{1}{2}\left(\ket{\Phi_{12}^+}P(\tau_{23})I + \ket{\Phi_{12}^-}P(\tau_{23})Z + \ket{\Psi_{12}^+}P(\tau_{23})X-i\ket{\Psi_{12}^-}P(\tau_{23})Y\right)\ket{\Psi_3^\text{in}}
	\end{split}
\end{equation}
with the standard Bell states $\ket{\Phi^\pm}$, $\ket{\Psi^\pm}$. The final representation uses the Pauli gates $I,X,Y,Z$ and a time-dependent phase gate $P(\tau_{23})$ 
\begin{equation}
	 \quad P(\tau_{23}) = \left[\begin{array}{cc}
			1&0\\0&\text{e}^{i\frac{FSS}{\hbar}\tau_{23}}
		\end{array}\right].
\end{equation}
Thus, the initial phase of the entangled photon pair is fully transferred onto the teleported state of photon 3. This does not directly reduce the faithfulness of the quantum teleportation process but leads, in turn, to a time-dependence of the final photon. The precise transformation between the input state $\rho_1^\text{in}$ and output state $\mathcal{E}(\rho_1^\text{in})$ can be described with the help of a 4x4 process matrix $\chi_{mn}$ via~\cite{Nielsen2012}
\begin{equation}
	\mathcal{E}(\rho_1^\text{in}) = \sum_{mn} \tilde{E}_m\rho_1^\text{in} \tilde{E}_n^\dagger\chi_{mn}
\end{equation}
with $\tilde{E}_i \in \{I,X,Y,Z\}$. In our experimental realization, we can correctly identify the Bell states $\ket{\Psi^\pm}$. In this case, the $\chi$-matrix of the teleportation process with the time-dependent entangled state reads (see supplementary material for further calculation)
\begin{equation}\label{eq:chi_time}
	\begin{split}
		\chi_{\ket{\Psi^\pm}}^\text{QT}(\tau_{23}) = \frac{1}{2}\left[\begin{array}{cccc}
			0 & 0 & 0 & 0\\
			0 & 1\pm\cos\left(\frac{FSS}{\hbar}\tau_{23}\right) & \pm\sin\left(\frac{FSS}{\hbar}\tau_{23}\right) &0 \\
			0 & \pm\sin\left(\frac{FSS}{\hbar}\tau_{23}\right) & 1 \mp \cos\left(\frac{FSS}{\hbar}\tau_{23}\right) & 0\\
			0 & 0 & 0 & 0
		\end{array}\right].
	\end{split}
\end{equation}
The time dependence of the entangled state is, in turn, reflected in the process matrices. For $\tau_{23} = 0$, the process matrices correspond to the ideal $\chi$-matrices for a time-independent entangled state, which we denote  as $\chi_{\ket{\Psi^\pm}}^\text{ideal} = \chi_{\ket{\Psi^\pm}}^\text{QT}(0)$. After $\tau_{23} \frac{FSS}{\hbar} = \pi$, however, the phase gate $P(\tau_{23})$ corresponds exactly to a Z-gate. Consequently, the process matrix is then also flipped with respect to the detection of the $\ket{\Psi}$-states with  $\chi_{\ket{\Psi^\pm}}^\text{QT}(\pi\frac{\hbar}{FSS}) = \chi_{\ket{\Psi^\mp}}^\text{ideal}$. Overall, an oscillatory behavior between $\chi_{\ket{\Psi^\pm}}^\text{ideal}$ is obtained according to the $FFS$-induced time evolution.\\
The time dependence of the teleportation process has an important consequence. For truthful information transfer in quantum teleportation, it is inherently required to apply a unitary operation on the teleported qubit. In this case, the unitary that must  applied changes over time. If the data are integrated agnostically to the underlying temporal dependence and then the unitary operation is applied, the fidelity of information transfer rapidly decreases for increasing time windows. However, if the underlying temporal structure is known, a drastic improvement of the fidelity can be achieved by applying the expected unitary operation for each time in the protocol. As we show later experimentally, a considerable improvement can already be achieved with restriction to pure Pauli gates. Any Pauli gate can be employed by simple basis flips in the respective other two bases which can be achieved without any experimental overhead.

\section{Results}
\begin{figure}[H]
	\centering
	\makebox[\textwidth][c]{\includegraphics{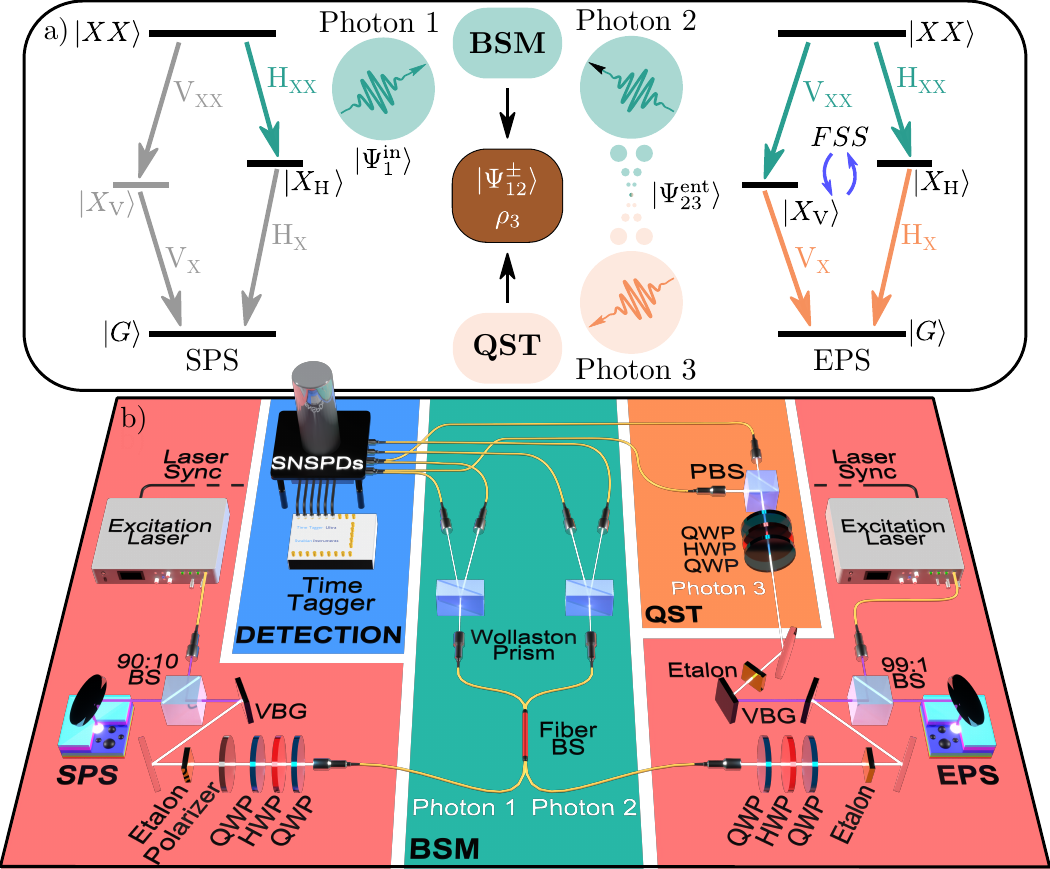}}
	\caption{a) Schematic of the full-photonic quantum teleportation implementation. Two semiconductors QDs are employed as single-photon source (SPS) and entangled photon pair source (EPS) utilizing the biexciton-exciton-cascade. Photon 1 (biexciton photon of the SPS) is prepared in a set of input states $\ket{\Psi_1^\text{in}}$ and interferes with photon 2 (biexciton photon of the EPS). A Bell state measurement reveals detection of the Bell states $\ket{\Psi_{12}^\pm}$ of the photons 1,~2. The polarization state $\rho_3$ of photon 3 (exciton photon of the EPS) is resolved via quantum state tomography (QST). Together, this allows to perform quantum process tomography (QPT) which enables the reconstruction of the process matrix of the quantum teleportation process. b) Experimental realization. SPS and EPS are located in two remote laboratory rooms connected via a \SI{50}{\meter} fiber link. Both sources are excited with two separate, synchronized lasers via resonant two-photon excitation. All three photons are individually filtered using volume Bragg grating (VBG) filters and etalons. A polarizer and a subsequent set of quarter-half-quarter waveplates (QWP-HWP-QWP) prepare the input states $\ket{\Psi_1^\text{in}}$ of photon 1. For the photons 2,~3, each a set of QWP-HWP-QWP matches the EPS polarization eigenbasis to the laboratory frame. The polarization state $\rho_3$ of photon 3 is analyzed using the same set of waveplates realizing a QST setup. Time-resolved characterization of the teleportation process is achieved with superconducting nanowire single-photon detectors (SNSPDs) and a TimeTagger which yield a total detection time jitter of about \SI{60}{\pico\second} (between two photons).}
	\label{fig:Setup}
\end{figure}

To generate the three photons, we operate a single-photon source (SPS) and an entangled photon pair source (EPS) according to the schematic and experimental realization depicted in figure \ref{fig:Setup} a) and b), respectively. The EPS is located in a separate laboratory room in a transportable, rack-integrated setup~\cite{Vyvlecka2026} and is connected to the SPS via a \SI{50}{\meter} fiber link. Both sources share an identical design and are based on InAs QDs emitting directly in the telecom C-band. For an increased efficiency of light extraction, the QD layer is incorporated into a planar cavity structure based on AlAs/GaAs and SiO$_2$/TiO$_2$ distributed Bragg reflectors~\cite{Joos2026}. Both sources are operated at about \SI{4}{\kelvin} in cryostats and we employ resonant two-photon excitation (TPE) in order to coherently populate the biexciton state. We use two distinct synchronized pulsed excitation lasers with a repetition frequency of \SI{76}{\mega\hertz} to precisely match the respective spectral TPE resonances. After collection, the excitation laser is filtered via volume Bragg grating (VBG) filters. For the EPS~\cite{Joos2026}, biexciton (photon 2) and exciton (photon 3) are separated via VBG filters and pass through an etalon with \SI{17}{\giga\hertz} bandwidth. For the SPS, we filter the biexciton (photon 1) emission with a VBG and the same type of etalon and utilize a linear polarizer followed by waveplates in order to prepare the different input states $\ket{\Psi_1^\text{in}}$ for the teleportation experiment. Photons 1 and 2 meet in the Bell state measurement (BSM) setup with the same wavelength of \SI{1537.14}{\nano\meter}. To achieve this common wavelength, a piezo-electric actuator beneath the sample membrane of the SPS is used, overcoming the initial offset of about \SI{20}{\giga\hertz}. The BSM setup allows for the successful identification of $\ket{\Psi^\pm}$ states whereas $\ket{\Phi^\pm}$ can not be distinguished. Upon measurement of a Bell state, the polarization state $\rho_3$ of photon 3 is investigated via quantum state tomography (QST).

\begin{figure}[H]
	\centering
	\makebox[\textwidth][c]{\includegraphics{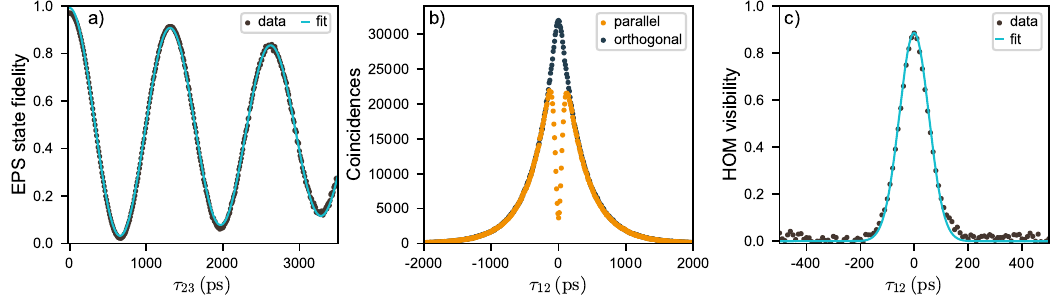}}
	\caption{Characterization of the sources. a) Time-resolved EPS state fidelity $\mathcal{F}(\rho_\text{EPS}, \ket{\Phi^+})$. b) HOM interference measurements between photon 1 and photon 2 for parallel and orthogonal polarization setting. c) Calculated HOM visibility $V$ between photon 1 and photon 2 dependent on $\tau_{12}$.}
	\label{fig:Sources}
\end{figure}
Before performing the teleportation experiment, we characterize the two employed non-classical photon sources. This will later enable a precise description of the time dependence of the teleportation process given the certain imperfections of the two sources. The two decisive parameters for faithful quantum teleportation are the degree of entanglement of the photon pair 2,~3 as well as the indistinguishability between photons 1 and 2. First, we perform QST of the polarization-entangled two-photon state emitted by the EPS and reconstruct the density matrix $\rho_\text{EPS}$ by means of maximum likelihood estimation (MLE). Figure \ref{fig:Sources} a) shows the measured fidelity $\mathcal{F}(\rho_\text{EPS}, \ket{\Phi^+})$ dependent on the temporal photon separation $\tau_{23}$. At zero time delay, a maximum fidelity of \SI{96.95\pm0.04}{\percent} confirms state-of-the-art performance for telecom C-band entangled photon pair sources~\cite{Joos2026}. Due to the non-zero $FFS$, an oscillatory behavior is observed. A fit to the data (see supplementary material for the fit model) allows the determination of the $FSS$ as well as a characterization of the dephasing mechanisms of the exciton spin-state~\cite{Hudson2007}, as summarized in table \ref{tab:SourceSummary}. Furthermore, we measure the two-photon interference between photons 1 and 2 in a Hong-Ou-Mandel interferometer. Figure \ref{fig:Sources} b) shows the correlations for parallel and orthogonal input polarizations. A strong dip around zero time delay can be observed for the parallel measurement. However, spectral broadening of both emitters affects the interference for larger time delays. Similarly to the fidelity, we calculate the HOM visibility dependent on the temporal photon separation $\tau_{12}$ (see figure \ref{fig:Sources} c)) as $V(\tau_{12}) = 1-C_\parallel(\tau_{12})/C_\perp(\tau_{12})$ using the time-resolved coincidences $C_\parallel$ ($C_\perp$) for the parallel (orthogonal) measurement. For small time delays, high visibility values up to \SI{88.5\pm0.3}{\percent} are achieved. A fit to the data (see supplementary material for the fit model) yields the broadening parameter $\Gamma = \sqrt{\Gamma_\text{P1}^2 + \Gamma_\text{P2}^2} = \SI{7.72\pm0.07}{\giga\hertz}$ (here in full-width at half-maximum of natural frequency) governed by the full inhomogeneous linewidth $\Gamma_{\text{P}i}$ of the interfering photons of both emitters~\cite{Kambs2018}. Furthermore, the fit reveals that the gap between the maximum visibility and unity is solely due to the finite temporal resolution of the detection system of \SI{60}{\pico\second}. Additional characterization measurements of the sources are provided in the supplementary material.

\begin{figure}[H]
	\centering
	\makebox[\textwidth][c]{\includegraphics{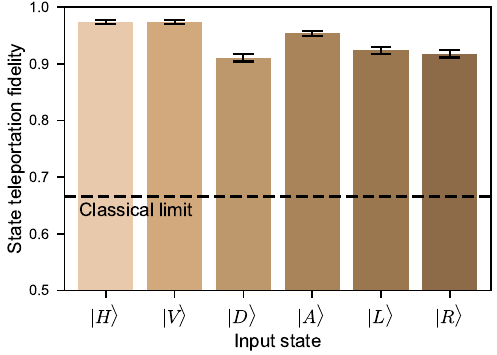}}
	\caption{State teleportation fidelity $\mathcal{F}(\rho_3, Y\ket{\Psi_1^\text{in}})$ at $\tau_{12} = \tau_{23} = 0$ (\SI{30}{\pico\second} time window) between input and output states for all six mutually unbiased input states. The average fidelity across the six states is \SI{94.2\pm0.5}{\percent}.}
	\label{fig:Output_states}
\end{figure}

\begin{table}
	\centering
	\caption{Summary of the values describing the performance of the EPS as well as the two-photon interference between the two sources as obtained from the measurements in figure \ref{fig:Sources}. Additionally, the lifetimes of three photons (see supplementary material and~\cite{Joos2026a}) as well as the average count rates during the measurement are stated.}
	\label{tab:SourceSummary}
	\begin{tabular}{lr}
		\toprule
		\textbf{Quantity} & \textbf{Value}\\
		\midrule
		Maximum EPS state fidelity - $\mathcal{F}(\rho_\text{EPS}, \ket{\Phi^+})$ & \SI{96.95\pm0.04}{\percent}\\
		Fine-structure splitting - $FFS$ (fit) & \SI{3.150\pm0.001}{\micro\electronvolt}\\
		Oscillation period - $T_\text{FFS}$ (fit) & \SI{1313\pm1}{\pico\second}\\
		Spin-scattering - $T_\text{SS}$ (fit) & \SI{18.6\pm0.3}{\nano\second}\\
		Cross-dephasing of EPS exciton state - $T_\text{HV}$ (fit) & \SI{17.6\pm0.3}{\nano\second}\\
		\midrule
		Maximum HOM visibility - $V$ & \SI{88.5\pm0.3}{\percent}\\
		Broadening parameter - $\Gamma$ (fit) & \SI{7.72\pm0.07}{\giga\hertz}\\
		\midrule
		Lifetime photon 1 (SPS, biexciton) - $T_\text{P1}$ & \SI{374\pm1}{\pico\second}\\
		Lifetime photon 2 (EPS, biexciton) - $T_\text{P2}$ & \SI{294\pm1}{\pico\second}\\
		Lifetime photon 3 (EPS, exciton) - $T_\text{P3}$ & \SI{757\pm1}{\pico\second}\\
		\midrule
		Count rate photon 1 (SPS, biexciton) -  $C_\text{P1}$ & \SI{573}{\kilo\counts}\\
		Count rate photon 2 (EPS, biexciton) -  $C_\text{P2}$ & \SI{3465}{\kilo\counts}\\
		Count rate photon 3 (EPS, exciton) -  $C_\text{P3}$ & \SI{2042}{\kilo\counts}\\
	\end{tabular}
\end{table}
For a precise description of the teleportation process, we perform quantum process tomography (QPT)~\cite{Chuang1997, Nielsen1998}. For this purpose, we prepare photon 1 in four polarization states: $\ket{H}$, $\ket{V}$, $\ket{D} = (\ket{H} + \ket{V})/\sqrt{2}$ and $\ket{L} = (\ket{H} + i\ket{V})/\sqrt{2}$. For each input state $\ket{\Psi_1^\text{in}}$, we measure the teleported polarization state $\rho_3$ of photon 3 via QST upon detection of a $\ket{\Psi^\pm_{12}}$ Bell state between photon 1 and photon 2. In the following, we discuss the teleportation results after detection of $\ket{\Psi^-_{12}}$. Detection of $\ket{\Psi^+_{12}}$ yields qualitatively the same results which are presented in the supplementary material. Finally, the $\chi$-matrix is obtained from the coincidence statistics via MLE~\cite{Knee2018}. Error propagation of the stated fidelities is performed via 1000 iterations of Monte-Carlo-simulations based on the measured coincidence statistics. In order to investigate the full time-evolution of the teleportation process, we determine the $\chi$-matrix dependent on the detection times of the three photons. By employing one channel of the BSM as a common start channel, we obtain two time delays. Firstly, the time delay to the detection of the second photon in the BSM which corresponds to $\tau_{12}$. Secondly, the time delay $\tau_\text{Tomo-BSM}$ to the detection of photon 3 in the QST setup. If photon 2 is detected on the BSM start channel, then $\tau_\text{Tomo-BSM}$ corresponds to $\tau_{23}$. Otherwise, if photon 1 yields the click on the BSM start channel, $\tau_{23}$ is the time difference to the BSM stop channel, which equivalently means $\tau_{23} = \tau_\text{Tomo-BSM} - \tau_{12}$. Previous demonstrations of quantum teleportation with QD photons~\cite{Reindl2018,BassoBasset2021, Strobel2025, Laneve2025} investigated the average teleportation fidelity for different sizes of integration time windows. However, the teleportation fidelity strongly differs along the two time axes.
\begin{figure}[H]
	\centering
	\makebox[\textwidth][c]{\includegraphics{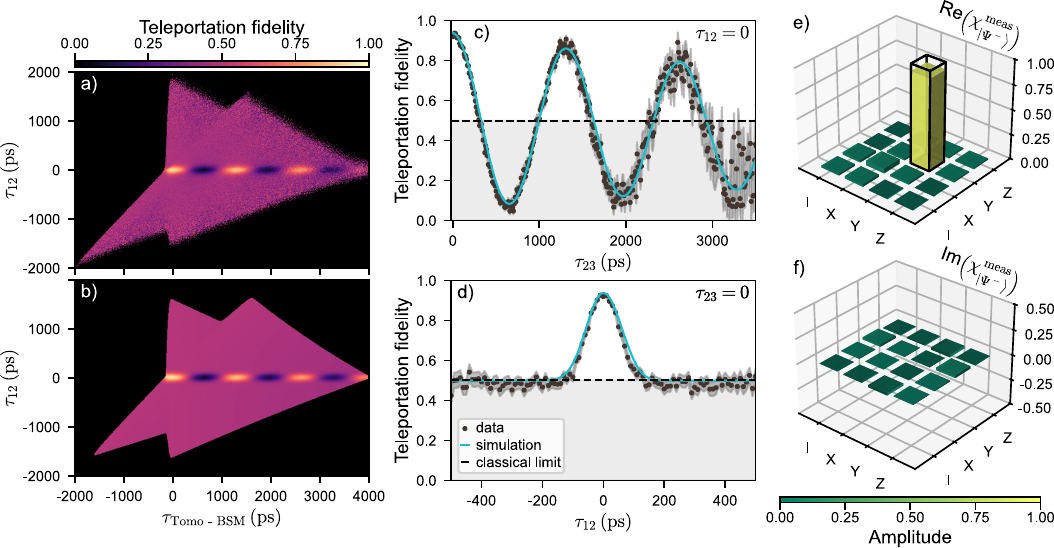}}
	\caption{Quantum teleportation with detection of $\ket{\Psi^-_{12}}$. a) Time-resolved teleportation fidelity $\mathcal{F}\left(\chi_{\ket{\Psi^-}}^\text{meas},\chi_{\ket{\Psi^-}}^\text{ideal}\right)$ of the measured $\chi_{\ket{\Psi^-}}^\text{meas}$ towards the static process matrix $\chi_{\ket{\Psi^-}}^\text{ideal}$). b) Simulated fidelity $\mathcal{F}\left(\chi_{\ket{\Psi^-}}^\text{simu},\chi_{\ket{\Psi^-}}^\text{ideal}\right)$ based on the characterization of the sources. c), d) Evolution of $\mathcal{F}\left(\chi_{\ket{\Psi^-}}^\text{meas},\chi_{\ket{\Psi^-}}^\text{ideal}\right)$ dependent on $\tau_{23}$ (time evolution  of the entanglement of the EPS) and $\tau_{12}$ (change of interference visibility between photon 1 and photon 2), respectively. e), Real and f) imaginary part of the measured process matrix $\chi_{\ket{\Psi^-}}^\text{meas}$ at $\tau_{12} = \tau_{23} = 0$ for a \SI{30}{\pico\second} time window. The fidelity to the ideal process matrix $\chi_{\ket{\Psi^-}}^\text{ideal}$ (black frames) is \SI{92.7\pm0.5}{\percent}.}
	\label{fig:Process_matrix}
\end{figure}
This can be seen in figure \ref{fig:Process_matrix} a) which presents a detailed view of the full time dependence showing the time-resolved teleportation fidelity $\mathcal{F}\left(\chi_{\ket{\Psi^-}}^\text{meas},\chi_{\ket{\Psi^-}}^\text{ideal}\right)$ of the measured $\chi$-matrix to $\chi_{\ket{\Psi^-}}^\text{ideal}$. For optimal interference visibility around $\tau_{12} = 0$, the fidelity oscillates due to the time-evolution of the entangled state $\ket{\Psi_{23}^\text{ent}(\tau_{23})} = \frac{1}{\sqrt{2}}\left(\ket{H_2H_3}+\text{e}^{i\frac{FSS}{\hbar}\tau_{23}}\ket{V_2V_3}\right)$. However, this oscillation is significantly reduced as $\tau_{12}$ increases due to the decreasing interference visibility between the photons 1 and 2. The black area corresponds to data points where the overall coincidence level is below 0.5\% of the maximum at the (0,0)-time coordinate, and no meaningful $\chi$-matrix can be reconstructed. The overall kite-like structure stems from the discussed relation between $\tau_\text{Tomo-BSM}$ and $\tau_{23}$. Further discussion about the time structure of the data is provided in the supplementary material. Based on the time dependence of interference visibility $V(\tau_{12})$, spin-flip probability $P_\text{SS}(\tau_{23})$, and cross-dephasing probability $P_\text{HV}(\tau_{23})$ (see table \ref{tab:SourceSummary}), we calculate the complementary theoretical process matrix (time-dependence is omitted for improved visual clarity; see supplementary material for detailed derivation)
\begin{equation}\label{eq:x_real}
	\begin{split}
		\chi^\text{simu}_{\ket{\Psi^-}} = V(1-P_\text{SS})(1-P_\text{HV})\chi_{\ket{\psi^-}}^\text{QT} + (1-P_\text{SS})(VP_\text{HV} + 1-V)\chi^\text{class} + P_\text{SS}\chi^\text{mix}.
	\end{split}
\end{equation}
Three components determine the real process matrix with source imperfections~\cite{BassoBasset2021}. First, the time-dependent process matrix $\chi_{\ket{\psi^-}}^\text{QT}$ for quantum teleportation, as described in equation \eqref{eq:chi_time}, which is obtained for successful interference and no spin-dephasing effects. If either cross-dephasing occurs or the interference fails, classical correlations $\chi^\text{class}$ are still present due to measurements in the eigenbasis of the entangled state. Finally, spin-scattering fully scrambles the polarization state, leading to a mixed process $\chi^\text{mix}$. Based on equation \eqref{eq:x_real}, we model the expected fidelity, as illustrated in figure \ref{fig:Process_matrix} b). Owing to the diligent source characterization, the modeled fidelity accurately matches the measured data. In order to gain deeper insight into the time dependence, we depict the temporal evolution of the fidelity of the data and simulation dependent on the time delays $\tau_{23}$ and $\tau_{12}$ in figure  \ref{fig:Process_matrix} c) and d), respectively. As anticipated, the fidelity oscillates with increasing $\tau_{23}$ due to the $FSS$-induced phase evolution. Thereby, the quantum process changes between $Y$- and $X$-gate. A minor reduction of the teleportation fidelity for larger time delays originates from the increased spin-scattering and cross-dephasing probability. On the other hand, increasing the photon separation $\tau_{12}$ is accompanied with a decrease in interference visibility. Because the classical correlations are preserved, a statistical mixture between $Y$ and $X$ processes is obtained. Consequently, the fidelity converges to 0.5 which is the limit for fidelity towards the ideal quantum process matrix achievable with classical methods~\cite{Horodecki1999, Nielsen2002,Olmschenk2009,Ma2012}. For vanishing time delays ($\tau_{12} = \tau_{23} = 0$, \SI{30}{\pico\second} time window), the measured process matrix fidelity reaches a value of \SI{92.7\pm0.5}{\percent}. The real and imaginary parts of corresponding process matrix are depicted in figure \ref{fig:Process_matrix} e), f), respectively. Another commonly used metric is the average fidelity between input and output states. For this, we prepared additionally the input states $\ket{A} = (\ket{H} - \ket{V})/\sqrt{2}$ and $\ket{R} = (\ket{H} - i\ket{V})/\sqrt{2}$ and measure the teleported state of photon $\rho_3$ for all six mutually unbiased input states. The resulting state teleportation fidelity $\mathcal{F}(\rho_3, Y\ket{\Psi_1^\text{in}})$ is depicted in figure \ref{fig:Output_states} ($\tau_{12} = \tau_{23} = 0$, \SI{30}{\pico\second} time window). High state teleportation fidelities of $>\SI{90}{\percent}$ are obtained for all six input states whereas values of about \SI{97}{\percent} are achieved for the $\ket{H}$ and $\ket{V}$ input states which don't require interference for successful teleportation. The average state teleportation fidelity across the six input states amounts to \SI{94.2\pm0.5}{\percent} far above the classical limit of 2/3~\cite{Massar1995, Horodecki1999, Nielsen2002, Olmschenk2009}. These high process matrix and average state teleportation fidelities surpass previous implementations of full photonic quantum teleportation with quantum emitters~\cite{Strobel2025, BassoBasset2021,Laneve2025}. Furthermore, even the performance of other (partially) photonic teleportation schemes with atoms~\cite{Nolleke2013, Olmschenk2009}, defect-centers in diamond~\cite{Ito2025} or spontaneous-parametric down-conversion sources~\cite{Akin2025, Peng2025} is matched for our peak fidelity.

Given the established understanding of the temporal evolution of the teleportation process, this can be harnessed to optimize the rate of faithful information transfer. In principle, it would be beneficial to increase the time window to consider all detected coincidences. However, this is connected to a reduction of the teleportation fidelity due to the source imperfections. Therefore, it is wise to consider the different time dependence along the two time axes for an optimized operation of quantum communication protocols. Here, we investigate exemplarily the dependence of the coincidence rate and teleportation fidelity on three different selections of time windows. 1) Firstly, a time window which is linearly increasing along both time axes, i.e. agnostic towards the underlying time evolution~\cite{Strobel2025, Laneve2025}. 2) Secondly, we calculate the optimal selection of time bins that maximize the simulated teleportation fidelity for a given total window size (area of the 2D time space). In this way, the effect of the rapidly decreasing interference visibility for increasing $\tau_{12}$ can be diminished. 3) Finally, we perform the same optimization as in step two. However, additionally the application of Pauli gates on the raw data is included. This is performed by bases flips on the raw data (in this case it is only necessary in the $X$ and $Y$ measurement bases) and can be implemented without any experimental overhead. This allows the achievement of high teleportation fidelity for the initial minima in the fidelity towards the static process matrix $\chi_{\ket{\Psi^-}}^\text{ideal}$ (see figure \ref{fig:Process_matrix} a), c)) reducing the negative impact of the QD $FSS$. A visualization of the time windows for the three scenarios can be found in the supplementary material. Figure \ref{fig:Fidelity_increasing} shows the resulting combination of the teleportation fidelity and three-photon coincidence rate for the three different scenarios. For small time windows, similar values are obtained for all methods. However, as a larger share of the maximum coincidence rate (\SI{320}{\counts} for $\ket{\Psi^-_{12}}$ detection) is used, a significant advantage of methods 2) and 3) is apparent. For instance a fidelity of \SI{80}{\percent} is accompanied by a coincidence rates of \SI{12.0}{\counts} with method 1) while employing method 3) boosts this number to \SI{25.9}{\counts}. This coincidence rate constitutes an improvement of more than two orders of magnitude compared to previous implementations of full-photonic quantum teleportation with remote emitters at comparable teleportation fidelity~\cite{Laneve2025}.\\
In order to achieve a higher teleportation fidelity for larger time windows, the two important source parameters, two-photon interference visibility between the remote sources and EPS state fidelity, must be improved for larger photon separation. Here, the interference visibility is the main limitation to the performance of the teleportation protocol. The incorporation of QD source into diode structures has shown to strongly decrease the linewidth broadening of remote emitters~\cite{Zhai2022}. On the other hand, the non-zero $FSS$ still leads to a reduction of the teleportation fidelity even when applying method 3). In this context, multi-axial piezo-electric actuators can be used to eliminate any residual asymmetry of the QD confinement~\cite{Trotta2016}. Consequently, the high maximum degree of entanglement achieved here can be preserved for the full wavepackage and the multi-axial piezo can be simultaneously used to achieve the wavelength matching of the two sources. Importantly, both techniques are fully compatible with the source design employed in this work.
\begin{figure}[H]
	\centering
	\makebox[\textwidth][c]{\includegraphics{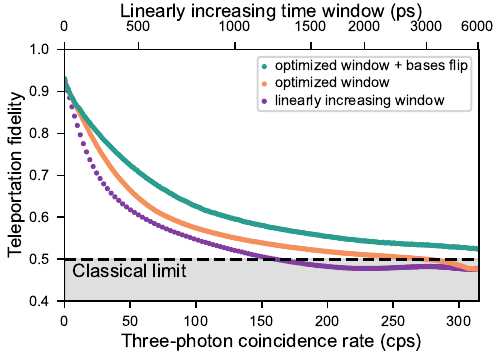}}
	\caption{Measured time-integrated teleportation fidelity towards the static process matrix $\chi_{\ket{\Psi^-}}^\text{ideal}$) and corresponding three-photon coincidence rate for three different selections of time windows. Purple dots: Along both time axes linearly increasing window. Orange dots: Optimized selection of time bins based on the theoretical modeling. Turquoise dots: Optimized selection of time bins based on the theoretical modeling plus basis flip (for $X$/$Y$ bases of the raw data) when $\mathcal{F}\left(\chi_{\ket{\Psi^-}}^\text{simu},\chi_{\ket{\Psi^-}}^\text{ideal}\right)<\mathcal{F}\left(\chi_{\ket{\Psi^-}}^\text{simu},\chi_{\ket{\Psi^+}}^\text{ideal}\right)$.}
	\label{fig:Fidelity_increasing}
\end{figure}

\section{Conclusion}
We demonstrate the first implementation of quantum teleportation using remote quantum emitters directly at the telecom C-band. To enable this demonstration, the biexciton-exciton transitions of two semiconductor QDs are brought into resonance using a piezo-electric actuator enabling two-photon interference between photons from distinct QDs. The high three-photon coincidence rate of \SI{320}{\counts} (for $\ket{\Psi^-_{12}}$ detection) allows to study and fully understand the time-dependence of the quantum teleportation process with remote QDs and its current limitations given the source imperfections. At the same time, the performance of the sources facilitates record values of the quantum teleportation fidelity of up to \SI{92.7\pm0.5}{\percent} (towards the ideal quantum process matrix) and average state teleportation fidelity of up to \SI{94.2\pm0.5}{\percent} (for six mutually unbiased input states) for full-photonic quantum teleportation with quantum emitters. The main limitation of the implementation in this work is the two-photon interference visibility between the two sources. Integration of the QDs into diode structures~\cite{Zhai2022} could prospectively strongly improve the performance for larger coincidence windows. Given the understanding of the time dependence of the limitations of the teleportation process in the current implementation, we present an optimization of the classical data-processing, necessary for quantum teleportation protocols, to maximize the achievable teleportation fidelity and coincidence rate. This work opens the route towards the operation of large-scale fiber-based quantum networks utilizing remote sources in quantum teleportation~\cite{Ma2012} or quantum repeater~\cite{VanLoock2020, Azuma2023} protocols.

\begin{backmatter}
\bmsection{Funding}
The authors acknowledge funding  by the German Federal Ministry of Research, Technology and Space (BMFTR) via the projects QR.N (16KIS2207) and QCyber (16KIS2590K). Additional funding was also provided via the project EQSOTIC. This project was funded within the QuantERA II Programme that has received funding from the EU’s H2020 research and innovation programme under the GA No 101017733, and with funding organisation BMFTR (with project number 16KIS2060K).
	
\bmsection{Acknowledgment}
We thank Tobias-Huber Loyola and Sven Höfling for the fabrication of the dielectric mirrors on the samples. We would like to gratefully acknowledge the help of Armando Rastelli for providing the piezo-electric actuator for wavelength tuning of the SPS and Bettina Frank for support for the fabrication of sample markers. We would like to thank Alessandro Laneve for fruitful discussions on MLE of QPT data. We thank SingleQuantum for their persistent support with the SNSPDs. Assistance of the chat bot Claude Sonnet 5.5 was used for the generation of a Blender file that was used to render the setup illustration in figure \ref{fig:Setup} b).


\end{backmatter}
\bibliography{References}

\end{document}